\documentclass[letterpaper, 10 pt, conference]{ieeeconf}

\IEEEoverridecommandlockouts

\usepackage{cite}
\usepackage{amsmath,amssymb,amsfonts}
\usepackage{algorithmic}
\usepackage{graphicx}
\usepackage{textcomp}
\usepackage{xcolor}
\usepackage[switch]{lineno}
\usepackage{algorithm}

 \usepackage[italian]{babel}
 \usepackage{pstricks}
 \usepackage{verbatim}
 \usepackage{graphicx}
 \usepackage{psfrag}
 \usepackage{fancybox}
 \usepackage{amsmath}
 \usepackage{amsfonts}
 \usepackage{amssymb}
 \usepackage{listings}
 \usepackage{enumerate}
 \usepackage{url}
 \usepackage{orcidlink}

\newcommand\notsotiny{\@setfontsize\notsotiny\@vipt\@viipt}
 \makeatother

\input{pst-plot}
\input{pst-node}
\input{pst-coil}
 \psset{unit=1.0\unitlength}
 \SpecialCoor

\newcommand{\flowright}[1]{\begin{pspicture}(0,12)
   \put(0,-1){\makebox(0,0)[t]{$\Rightarrow$}}
\end{pspicture}}

\newcommand{\flowleft}[1]{\begin{pspicture}(0,12)
   \put(0,-1){\makebox(0,0)[t]{$\leftarrow$}}
\end{pspicture}}

\newcommand{\tras}{^{\mbox{\tiny T}}}

\newcommand{\muno}{^{\mbox{\tiny -1}}}

\newcommand{\ts}{\textstyle}
\newcommand{\scr}{\scriptstyle}

 \newcommand{\mat}[2]{\left[\begin{array}{#1} #2 \end{array}\right]}

  \def\G{{\bf G}} 
 \def\J{{\bf J}}  
\def\M{{\bf M}} \def\N{{\bf N}}  
 \def\R{{\bf R}}

 \newcommand{\Mca}{M_{ca}}
 \newcommand{\Mcb}{M_{cb}}
 \newcommand{\Mi}{M_{1}}
 \newcommand{\Mii}{M_{2}}

 \newcommand{\Ra}{R_a}
 \newcommand{\Rb}{R_b}

 \newcommand{\omegasq}{\omega^{\mbox{\tiny$\square$}}}
 \newcommand{\wsq}{\bomega^{\mbox{\tiny$\square$}}}
 \newcommand{\Deltaomega}{\delta}
 \newcommand{\Deltaw}{\boldsymbol{\delta}}
 \newcommand{\bomega}{\boldsymbol{\omega}}

\renewcommand{\keywords}[1]{\textbf{\textit{Index terms---}} #1}

\definecolor{mygreen}{RGB}{28,172,0}

\newcommand{\ENG}{1=1}                              

\newcommand{\ItaEng}[2]{\ifnum\ENG\relax#2\else#1\fi}

\definecolor{mio_red}{rgb}{0.0, 0.0, 0.0}
\definecolor{mioo_red}{rgb}{0.0, 0.0, 0.0}

\title{\LARGE \bf
Discrete-Time Model of a Two-Speed PowerShift suitable for Real-Time Control and Simulation
}

\author{Riccardo Morselli\orcidlink{0009-0009-2007-9282}, Davide Tebaldi\orcidlink{0000-0003-1432-0489} \IEEEmembership{Member, IEEE}  and Roberto Zanasi\orcidlink{0000-0001-5507-825X}
\thanks{The work was partly supported by the University of Modena and Reggio Emilia
through the action FARD (Finanziamento Ateneo Ricerca Dipartimentale) 2023/2024, and funded under the National Recovery and Resilience Plan (NRRP), Mission 04 Component 2 Investment 1.5 - NextGenerationEU, Call for tender n. 3277 dated 30/12/2021
Award Number:  0001052 dated 23/06/2022.}
\thanks{Riccardo Morselli is with Allison OH. E-mail: riccardo.morselli@alsn.com}
\thanks{Davide Tebaldi and Roberto Zanasi are
with the Department of Engineering ``Enzo Ferrari'', University of
Modena and Reggio Emilia, Modena, Via Pietro Vivarelli 10 - int. 1 -
41125 Modena, Italy. E-mails: davide.tebaldi@unimore.it, roberto.zanasi@unimore.it.}
}

\begin{document}

\maketitle
\thispagestyle{empty}
\pagestyle{empty}

\begin{abstract}

In this paper, a new discrete-time approach to model the clutches engagement/disengagement in a two-speed powershift is proposed. The core idea is the development of a model for the computation of the exact torque needed to achieve the clutches engagement, including both the cases of single clutch engagement and of simultaneous clutch engagement (full lock condition). Based on this, the control logic for the clutches engagement and disengagement phases is also developed. The advantages in terms of 
real-time applicability 
with respect to the continuous-time version are shown through extensive simulation results.

\end{abstract}


\section{Introduction}
Vehicle powertrains can be defined as complex physical systems
composed of several different physical subsystems interacting with each other through energetic ports~\cite{Miller2006}.

From the point of view of control engineers, the establishment of accurate and detailed models~\cite{bouscayrol2023energetic,Tebaldi2025} is essential for the development of effective control strategies~\cite{Borja2023}.  Indeed, the development of approaches to model physical systems in different energetic domains has been the subject of studies
 by many renowned scientists throughout history \cite{maxwell1861}. Vehicle powertrains contain a number of physical systems with a high level of model complexity belonging to different energetic domains, ranging from mechanical systems such as differentials~\cite{Zhao2018} and planetary gear sets~\cite{Lhomme2017}, to electronic systems such as power converters~\cite{Nielsen2025}, all the way to electromechanical systems such as electric machines~\cite{Ramones2022}.

From the modeling point of view, a particularly challenging mechanism to model is the involvement of static and coulomb frictions, which are indeed widely used in automotive mechanical systems to control the synchronization between two shafts or two axles. Their simulation does pose significant challenges because of the variable dynamic dimension nature of the resulting system~\cite{Zanasi2001,van2007introduction}. Static and coulomb frictions are in fact adopted to describe the engagement/disengagement of multiple clutches in powershift transmissions~\cite{morselli2006modelling}. However, because of the complex nature of this phenomenon, the clutches engagement/disengagement is sometimes either not modeled in detail~\cite{LI2019} or modeled using average models~\cite{WALKER2011}.
\begin{figure}[t]
 \centering
 \setlength{\unitlength}{4.9mm}
 \psset{unit=\unitlength}
 \SpecialCoor
 \newrgbcolor{darkgreen}{0 0.5 0}
 \begin{pspicture}(0,-4)(12,9.5)
 {\small
\rput(4,7.5){Clutch a}
\rput(4,6.75){$1^{st}$ Gear}
\rput(8,9){Clutch b}
\rput(8,8.275){$2^{nd}$ Gear}
}
 \rput(0,-10){
 \psline[linewidth=2.0pt]{-}(2,9)(12,9)
 \psline[linewidth=1pt]{|-|}(4,6)(4,12) %
 \psline[linewidth=1pt]{|-|}(8,7)(8,10.5)
 \psarc[linewidth=0.1pt,linestyle=solid]{<-}(1,14){0.6}{-80}{80}     
 \rput[b](1.15,14.7){$\omega_1,\,\Mi$}
 \rput[b](5,15.35){$\scr\Mca$}
 \rput[b](7,15.35){$\scr\Mcb$}
 \rput[rb](3.9,14.1){$\ts\Deltaomega_a$}
 \rput[lb](8.1,14.1){$\ts\Deltaomega_b$}
 \rput[l](10.25,14){$J_1$}
 \rput[r](3.75,13.00){$r_{a1}$}
 \rput[l](8.25,12.25){$r_{b1}$}
 }
  \psline[linewidth=2.0pt]{-}(0,4)(10,4)
  \psline[linewidth=1pt]{|-}(4,6)(4,4.25)(5,4.25)(5,5.00)
  \psline[linewidth=1pt]{-}(5,4.25)(5.3,4.25)(5.3,5.00)
  \psline[linewidth=1pt]{-}(5.5,4)(5.5,5.20)(4.8,5.20)(4.8,4.4)
  \psline[linewidth=1pt]{-}(5.15,5.20)(5.15,4.4)
  \psline[linewidth=1pt]{|-}(4,2.1)(4,3.75)(5,3.75)(5,3.0)
   \psline[linewidth=1pt]{-}(5,3.75)(5.3,3.75)(5.3,3)
   \psline[linewidth=1pt]{-}(5.5,4)(5.5,2.80)(4.8,2.80)(4.8,3.6)
   \psline[linewidth=1pt]{-}(5.15,2.80)(5.15,3.6)
  \psline[linewidth=1pt]{|-}(8,7.5)(8,4.25)(7,4.25)(7,5.00)
  \psline[linewidth=1pt]{-}(7,4.25)(6.7,4.25)(6.7,5.00)
  \psline[linewidth=1pt]{-}(6.5,4)(6.5,5.20)(7.2,5.20)(7.2,4.4)
  \psline[linewidth=1pt]{-}(6.85,5.20)(6.85,4.4)
   \psline[linewidth=1pt]{|-}(8,0.6)(8,3.75)(7,3.75)(7,3.0)
  \psline[linewidth=1pt]{-}(7,3.75)(6.7,3.75)(6.7,3.00)
  \psline[linewidth=1pt]{-}(6.5,4)(6.5,2.80)(7.2,2.80)(7.2,3.6)
  \psline[linewidth=1pt]{-}(6.85,2.80)(6.85,3.6)
 \psarc[linewidth=0.1pt,linestyle=solid]{<-}(10,-1){0.6}{-80}{80}
 \rput[t](10.15,-1.7){$\omega_2,\,\Mii$}
 \rput[r](1.75,-1){$J_2$}
 \rput[l](8.25,-0.25){$r_{b2}$}
  \rput[r](3.75,0.5){$r_{a2}$}
\end{pspicture}
\vspace{-2mm}
\caption{Schematic representation of the considered two-speed powershift.}
\label{Sincronizzatore}
\vspace{-3.68mm}
\end{figure}
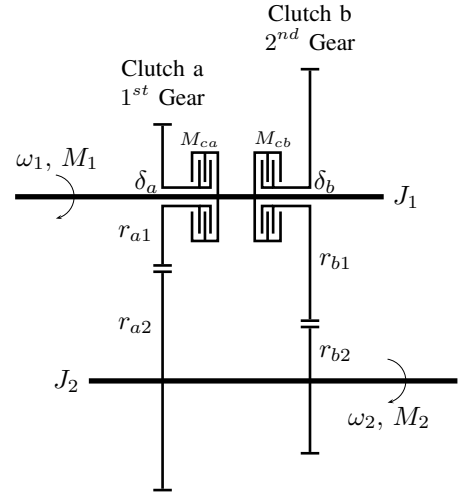

Past modeling solutions include the use of the LuGre dynamic friction model~\cite{Canudas1995}, which was later overcome by the approach proposed in~\cite{morselli2006modelling} in terms of both precision and computation time. However, the approach in~\cite{morselli2006modelling} still exhibits limitations due to its continuous-time nature, related to variable-step size that limits real-time application and the need for zero-crossing detection. To overcome these limitations, in this work we propose a discrete-time approach to compute the exact friction torques needed to achieve clutch engagement in a two-speed powershift. Based on this, the control logic for the clutches engagement and disengagement phases is also developed. To show the potentiality of the proposed approach, we perform extensive simulations by comparing the results with those offered by the continuous-time counterpart, and we comment them in terms of real-time applicability and computation time.

While this paper proposes a possible application of the introduced modeling approach, it is worth noting that this work paves the way for several other applications, such as the enhanced modeling of torque converters, brakes, and other complex devices typically found in vehicle powertrains.

The remainder of this paper is structured as follows. Sec.~\ref{Two_Speed_PowerShift_sect} addresses the proposed discrete-time modeling of the two-speed powershift, by also reporting the continuous-time counterpart. Sec.~\ref{control_sect} addressed the developed logic for the clutches engagement/disengagement, while Sec.~\ref{Simulations_results} presents the simulations results and the discussion. Finally, the conclusions of this work are drawn in Sec.~\ref{conclusions_sect}.

\section{Two-Speed PowerShift}\label{Two_Speed_PowerShift_sect}
The proposed discrete-time modeling of the
two-speed powershift is addressed in Sec.~\ref{Discrete_modeling_sect},  while the continuous-time
counterpart is addressed in Sec.~\ref{c_Cont_modeling_sect}. For both cases, the single gear engagement and full lock condition phases are modeled and commented in detail.

\subsection{Discrete-Time Modeling of the Two-Speed PowerShift}\label{Discrete_modeling_sect}

Reference is made to the schematic representation of the considered two-speed powershift shown in Fig.~\ref{Sincronizzatore}.
Let 
$
\Ra = r_{a2}/r_{a1}$ and
$\Rb = r_{b2}/r_{b1}$
denote the gear ratios associated with the first and second gears, that are engaged when clutch ``a'' and clutch ``b'' are closed, respectively.
Let the discrete velocity variables $\Deltaomega_a(k)$ and  $\Deltaomega_b(k)$ be defined as follows:
\begin{equation} \label{MA_eq_2}
\begin{array}{rcl}
\Deltaomega_a(k)
&=&
\omega_1(k)  - \Ra\,\omega_2(k),
\\[1mm]
\Deltaomega_b(k)
&=&
\omega_1(k)  - \Rb\,\omega_2(k).
\end{array}
\end{equation}
Eqs.~\eqref{MA_eq_2} can also be rewritten in matrix form as follows:
\begin{equation} \label{MA_eq_2_Matrix_Form}
\underbrace{
\mat{@{\,}c@{\,}}{ \Deltaomega_a(k) \\ \Deltaomega_b(k)}
}_{\Deltaw(k)}
=
-
\underbrace{
\mat{@{\,}c@{\;}c@{\,}}{ -1 &  \Ra \\ -1  & \Rb}
}_{\R\tras}
\underbrace{
\mat{@{\,}c@{\,}}{ \omega_1(k) \\ \omega_2(k)}
}_{\bomega(k)},
\end{equation}
or, equivalently, as follows:
\begin{equation} \label{MA_eq_2_MF}
\Deltaw(k)
=
-\R\tras\bomega(k).
\end{equation}
The maximum possible values of the friction torques $\Mca(k)$ and $\Mcb(k)$ can be expressed as follows:
\begin{equation} \label{MA_Mca_Mcb}
\begin{array}{rcl}
\Mca(k)
&=&
K_{ca}(k)\, \mbox{sign}\!\left(\Deltaomega_a(k) \right),
\\[1mm]
\Mcb(k)
&=&
K_{cb}(k) \,\mbox{sign}\!\left(\Deltaomega_b(k) \right),
\end{array}
\end{equation}
where $K_{ca}(k)$ and $K_{cb}(k)$ represent the saturation of the friction torques.

Using the Euler backward discretization method, the discrete accelerations  $\omegasq_1(k)$ and $\omegasq_2(k)$ of the two inertias $J_1$ and $J_2$ in Fig.~\ref{Sincronizzatore} can be defined as follows:
\begin{equation} \label{MA_omegasq_k_def}
\ts
\omegasq_1(k)
= \frac{\omega_1(k)-\omega_1(k-1)}{T_s},
\hspace{5mm}
\omegasq_2(k)
= \frac{\omega_2(k)-\omega_2(k-1)}{T_s},
\end{equation}
where the two relations in \eqref{MA_omegasq_k_def} denote the incremental ratios approximating the discrete-time derivatives.

Using \eqref{MA_eq_2} and \eqref{MA_Mca_Mcb},
the discrete-time dynamic equations of the two-speed powershift shown in Fig.~\ref{Sincronizzatore} are the following:
\begin{equation} \label{MA_eq_1}
\begin{array}{r@{\,}c@{\,}l}
J_1\omegasq_1(k)
&=&
\Mi(k) - \Mca(k) - \Mcb(k),
\\[1mm]
J_2\omegasq_2(k)
&=&
\Mii(k)  + \Ra \Mca(k) +\Rb  \Mcb(k),
\end{array}
\end{equation}
where $\Mi(k)$ and $\Mii(k)$ denote the input torques applied to the inertias $J_1$ and $J_2$, respectively.
Eqs.~\eqref{MA_eq_1} can also be written as follows:
\begin{equation} \label{MA_Sys_Dyn}
\underbrace{
\mat{@{\,}c@{\,}c@{\,}}{ J_1 & 0 \\ 0  & J_2}
}_{\J}
\underbrace{
\mat{@{\,}c@{\,}}{ \omegasq_1(k) \\ \omegasq_2(k)}
}_{\wsq(k)}
\!=\!
\underbrace{
\mat{@{\,}c@{\,}}{ \Mi(k) \\ \Mii(k)}
}_{\M(k)}
\!+\!
\underbrace{
\mat{@{\,}c@{\,}c@{\,}}{ -1 & -1 \\ \Ra  & \Rb}
}_{\R}
\underbrace{
\mat{@{\,}c@{\,}}{ \Mca(k) \\ \Mcb(k)}
}_{\M_c(k)},
\end{equation}
 or, equivalently,  in matrix form as follows:
\begin{equation} \label{MA_Sys_Dyn_MF}
\J
\wsq(k)
=
\M(k)
+
\R
\M_c(k),
\end{equation}
representing the discrete-time model of the considered two-speed powershift.
Left-multiplying Eq.~\eqref{MA_Sys_Dyn_MF} by $\J\muno$ yields:
\begin{equation} \label{MA_wsq_k}
\wsq(k)
=
\J\muno\left[\M(k)
+
\R
\M_c(k) \right].
\end{equation}
From \eqref{MA_omegasq_k_def}, the vector $\wsq(k)$ can be defined as follows:
\begin{equation} \label{MA_wsq_k_def}
\wsq(k)
= \frac{\bomega(k)-\bomega(k-1)}{T_s}.
\end{equation}
Therefore, from \eqref{MA_wsq_k} and \eqref{MA_wsq_k_def}, it  follows that:
\begin{equation} \label{MA_w_k}
\bomega(k)
=
\bomega(k-1)
+
\J_s\!\left(\M(k)
+
\R
\M_c(k) \right),
\end{equation}
where $\J_s=T_s\J\muno$.

\subsubsection{Engagement of a single gear}

Let $\G_1$ and $\G_2$ be the following selection vectors:
$$
\G_1=\mat{@{\,}c@{\;\;}c@{\,}}{ 1 & 0},
\hspace{12MM}\
\G_2=\mat{@{\,}c@{\;\;}c@{\,}}{ 0 & 1}.
$$
From \eqref{MA_eq_2_MF} it follows that
the $i$-th gear, for $i\!\in\!\{1, \; 2\}$, is engaged if the following  condition is satisfied:
\begin{equation} \label{MA_gear_eng}
\G_i\Deltaw(k) = 0
\hspace{8mm}
\Leftrightarrow
\hspace{8mm}
\G_i\R\tras\bomega(k) = 0.
\end{equation}
Substituting \eqref{MA_wsq_k_def} in \eqref{MA_gear_eng} yields:
\begin{equation} \label{MA_gear_eng_2}
\G_i\R\tras\left[\bomega(k\!-\!1)
+
\J_s\M(k) \right]
+
\G_i\J_R
\M_c(k)=0,
\end{equation}
where matrix $\J_R$ is defined as follows:
\begin{equation} \label{MA_J_R}
\J_R
=\R\tras\J_s\R
=\R\tras T_s\J\muno\R
=
\mat{@{\,}c@{\;\;}c@{\,}}{ J_R^{11} & J_R^{12}  \\[1mm] J_R^{21}   & J_R^{22} }.
\end{equation}
From \eqref{MA_gear_eng_2}, \eqref{MA_J_R} and \eqref{MA_Sys_Dyn}, the following relation holds:
\begin{equation} \label{MA_gear_eng_4}
\underbrace{
\mat{@{\,}c@{\;\;}c@{\,}}{ J_R^{i1} & J_R^{i2}}
}_{\G_i\J_R}
\underbrace{
\mat{@{\,}c@{\,}}{ \Mca(k) \\ \Mcb(k)}
}_{\M_c(k)}
\!=\!
\underbrace{
-\G_i\R\tras\left[
\J_s\M(k)
\!+\!
\bomega(k\!-\!1)
\right]
}_{\N_i(k)},
\end{equation}
which can be rewritten as follows:
\begin{equation} \label{MA_gear_eng_5}
 J_R^{i1}
\Mca(k)
+
J_R^{i2}\Mcb(k)
=
\N_i(k).
\end{equation}
From \eqref{MA_gear_eng_5} it follows that, for $i\!=\!1$, the first gear is engaged if:
\begin{equation} \label{MA_first_gear_eng}
   \Mca(k)
=
\frac{\N_1(k)- J_R^{12}\Mcb(k)}{J_R^{11}}.
\end{equation}
Equivalently, from \eqref{MA_gear_eng_5} it follows that, for $i\!=\!2$, the second gear is engaged if:
\begin{equation} \label{MA_second_gear_eng}
   \Mcb(k)
=
\frac{\N_2(k)- J_R^{22}\Mca(k)}{J_R^{21}}.
\end{equation}
Using \eqref{MA_Mca_Mcb} and \eqref{MA_eq_2}, the two engagement conditions \eqref{MA_first_gear_eng} and \eqref{MA_second_gear_eng} can be rewritten as follows:
\[
\begin{array}{@{}r@{\,}c@{\,}l@{}}
   \Mca(k)
&=&
\frac{\N_1(k)- J_R^{12}K_{ca}(k)\, \mbox{\footnotesize sign}\left(\omega_1(k)  - \Ra\,\omega_2(k) \right)}{J_R^{11}},
\\[3mm]
   \Mcb(k)
&=&
\frac{\N_2(k)- J_R^{21}K_{cb}(k)\, \mbox{\footnotesize sign}\left(\omega_1(k)  - \Rb\,\omega_2(k) \right)}{J_R^{22}}.
\end{array}
\]

\subsubsection{Full lock condition}

The considered two-speed powershift reaches the full lock condition (i.e. $\omega_1\!=\!\omega_2\!=\!0$) if $\Deltaw(k)=0$ or, equivalently, if:
\begin{equation} \label{MA_eq_2_MF_Full_lock}
\R\tras\bomega(k)=0,
\end{equation}
where the latter condition is derived from  \eqref{MA_eq_2_MF}.
Substituting \eqref{MA_w_k} in \eqref{MA_eq_2_MF_Full_lock} yields:
\begin{equation} \label{MA_w_k_Full_lock}
\R\tras\left[\bomega(k-1)
+
\J_s\!\left(\M(k)
+
\R
\M_c(k) \right)\right]=0.
\end{equation}
Solving \eqref{MA_w_k_Full_lock} with respect to $\M_c(k)$ yields the following full lock condition:
\begin{equation} \label{MA_w_k_Full_lock_condition}
\M_c(k)
=
-(\R\tras\J_s\R)\muno\R\tras\left(
\J_s\M(k)
\!+\!
\bomega(k\!-\!1)
\right).
\end{equation}
Since matrix $\R$ is invertible and $\J_s=T_s\J\muno$, the full lock condition \eqref{MA_w_k_Full_lock_condition} can also be rewritten as follows:
\begin{equation} \label{MA_w_k_Full_lock_condition_2}
\M_c(k)
=
-\R\muno\left(
\M(k)
\!+\!
\frac{\J}{T_s}\bomega(k\!-\!1)
\right),
\end{equation}
or, equivalently, as follows:
\begin{equation} \label{MA_w_k_Full_lock_condition_2}
\mat{@{\,}c@{\,}}{ \Mca(k) \\[1mm] \Mcb(k)}
=
-\R\muno\left(
\mat{@{\,}c@{\,}}{ \Mi(k) \\ \Mii(k)}
\!+\!
\frac{1}{T_s}
\mat{@{\,}c@{\,}}{ J_1\omega_1(k\!-\!1) \\ J_2\omega_2(k\!-\!1)}
\right),
\end{equation}
where the latter condition is obtained using \eqref{MA_eq_2_Matrix_Form} and \eqref{MA_Sys_Dyn}.

\subsection{Continuous-Time Modeling of the Two-Speed PowerShif}\label{c_Cont_modeling_sect}

Let the continuous velocity variables $\Deltaomega_a(t)$ and  $\Deltaomega_b(t)$ be defined as follows:
\begin{equation} \label{c_MA_eq_2}
\begin{array}{rcl}
\Deltaomega_a(t)
&=&
\omega_1(t)  - \Ra\,\omega_2(t),
\\[1mm]
\Deltaomega_b(t)
&=&
\omega_1(t)  - \Rb\,\omega_2(t).
\end{array}
\end{equation}
Eqs.~\eqref{c_MA_eq_2} can  be rewritten in matrix form as:
\[
\underbrace{
\mat{@{\,}c@{\,}}{ \Deltaomega_a(t) \\ \Deltaomega_b(t)}
}_{\Deltaw(t)}
=
-
\underbrace{
\mat{@{\,}c@{\;}c@{\,}}{ -1 &  \Ra \\ -1  & \Rb}
}_{\R\tras}
\underbrace{
\mat{@{\,}c@{\,}}{ \omega_1(t) \\ \omega_2(t)}
}_{\bomega(t)},
\]
or, equivalently, as follows:
\begin{equation} \label{c_MA_eq_2_MF}
\Deltaw(t)
=
-\R\tras\bomega(t).
\end{equation}
The maximum possible values of the friction torques $\Mca(t)$ and $\Mcb(t)$ can be expressed as follows:
\begin{equation} \label{c_MA_Mca_Mcb}
\begin{array}{rcl}
\Mca(t)
&=&
K_{ca}(t) \,\mbox{sign}\!\left(\Deltaomega_a(t) \right),
\\[1mm]
\Mcb(t)
&=&
K_{cb}(t) \,\mbox{sign}\!\left(\Deltaomega_b(t) \right),
\end{array}
\end{equation}
where $K_{ca}(t)$ and $K_{cb}(t)$
represent the saturation of the friction torques.
Using \eqref{c_MA_eq_2} and \eqref{c_MA_Mca_Mcb},
the continuous-time dynamic equations of the two-speed powershift shown in Fig.~\ref{Sincronizzatore} are the following:
\begin{equation} \label{c_MA_eq_1}
\begin{array}{r@{\,}c@{\,}l}
J_1\dot\omega_1(t)
&=&
\Mi(t) - \Mca(t) - \Mcb(t),
\\[1mm]
J_2\dot\omega_2(t)
&=&
\Mii(t)  + \Ra \Mca(t) +\Rb  \Mcb(t).
\end{array}
\end{equation}
Eqs.~\eqref{c_MA_eq_1} can also be written as follows:
\[
\underbrace{
\mat{@{\,}c@{\,}c@{\,}}{ J_1 & 0 \\ 0  & J_2}
}_{\J}
\underbrace{
\mat{@{\,}c@{\,}}{ \dot\omega_1(t) \\ \dot\omega_2(t)}
}_{\dot\bomega(t)}
\!=\!
\underbrace{
\mat{@{\,}c@{\,}}{ \Mi(t) \\ \Mii(t)}
}_{\M(t)}
\!+\!
\underbrace{
\mat{@{\,}c@{\,}c@{\,}}{ -1 & -1 \\ \Ra  & \Rb}
}_{\R}
\underbrace{
\mat{@{\,}c@{\,}}{ \Mca(t) \\ \Mcb(t)}
}_{\M_c(t)},
\]
 or, equivalently,  in matrix form as follows:
\begin{equation} \label{c_MA_Sys_Dyn_MF}
\J
\dot\bomega(t)
=
\M(t)
+
\R
\M_c(t),
\end{equation}
representing the continuous-time model of the considered two-speed powershift.
Left-multiplying Eq.~\eqref{c_MA_Sys_Dyn_MF} by $\J\muno$ yields:
\begin{equation} \label{c_MA_wsq_k}
\dot\bomega(t)
=
\J\muno\left[\M(t)
+
\R \M_c(t) \right].
\end{equation}


\begin{figure}[t]
\centering
\includegraphics[clip,width=1\linewidth]{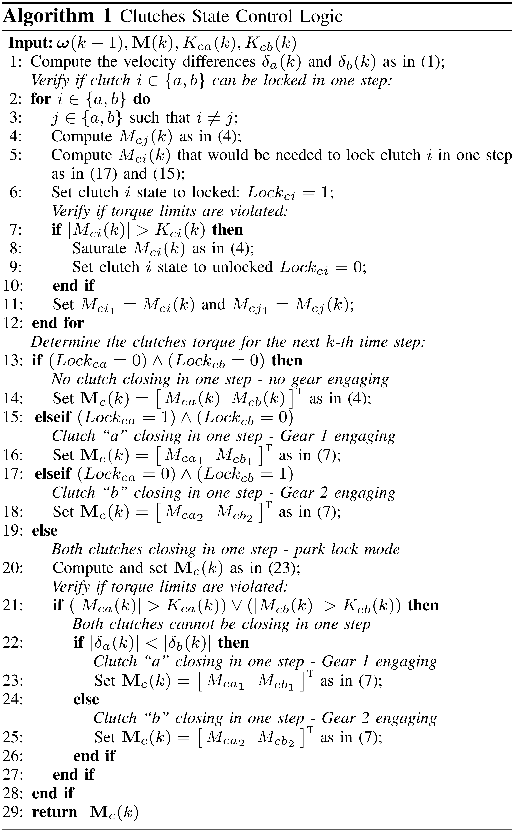}
 \end{figure}

\subsubsection{Engagement of a single gear}

From \eqref{c_MA_eq_2_MF}, it follows that
the $i$-th gear, for $i\!\in\!\{1, \; 2\}$, is engaged if the following  condition is satisfied:
\begin{equation} \label{c_MA_gear_eng}
\G_i\Deltaw(t) = 0
\hspace{8mm}
\Leftrightarrow
\hspace{8mm}
\G_i\R\tras\bomega(t) = 0.
\end{equation}
From \eqref{c_MA_gear_eng}, it also follows that:
\begin{equation} \label{c_MA_gear_eng_x}
\G_i\R\tras\dot\bomega(t) = 0.
\end{equation}
Replacing \eqref{c_MA_wsq_k} in \eqref{c_MA_gear_eng_x} yields
$\G_i\R\tras\J\muno\left[\M(t)
+
\R \M_c(t) \right]=0$,
from which it follows that:
\begin{equation} \label{c_MA_gear_eng_y}
\G_i
\underbrace{\R\tras\J\muno\R}_{\J_R} \M_c(t)
=
- \G_i\R\tras\J\muno \M(t).
\end{equation}
Eq.~\eqref{c_MA_gear_eng_y} can also be rewritten as:
\[
\underbrace{
\mat{@{\,}c@{\;\;}c@{\,}}{ J_R^{i1} & J_R^{i2}}
}_{\G_i\J_R}
\underbrace{
\mat{@{\,}c@{\,}}{ \Mca(t) \\ \Mcb(t)}
}_{\M_c(t)}
\!=\!
\underbrace{
-\G_i\R\tras
\J\muno\M(t)
}_{\N_i(t)},
\]
or, equivalently, as follows:
\begin{equation} \label{c_MA_gear_eng_5}
 J_R^{i1}
\Mca(t)
+
J_R^{i2}\Mcb(t)
=
\N_i(t).
\end{equation}
From \eqref{c_MA_gear_eng_5}, the condition giving the engagement of the first gear can be determined, namely if:
\begin{equation} \label{c_MA_first_gear_eng}
   \Mca(t)
=
\frac{\N_1(t)- J_R^{12}\Mcb(t)}{J_R^{11}}.
\end{equation}
Equivalently, the condition giving the engagement of the second gear can also be determined from \eqref{c_MA_gear_eng_5}, namely if:
\begin{equation} \label{c_MA_second_gear_eng}
   \Mcb(t)
=
\frac{\N_2(t)- J_R^{21}\Mca(t)}{J_R^{22}}.
\end{equation}
Using \eqref{c_MA_Mca_Mcb} and \eqref{c_MA_eq_2}, the two engagement conditions \eqref{c_MA_first_gear_eng} and \eqref{c_MA_second_gear_eng} can be rewritten as follows:
\begin{equation} \label{c_MA_first_and_second_gear_eng}
\begin{array}{@{}r@{\,}c@{\,}l@{}}
   \Mca(t)
&=&
\frac{\N_1(t)- J_R^{12}K_{ca}(t)\, \mbox{\footnotesize sign}\left(\omega_1(t)  - \Ra\,\omega_2(t) \right)}{J_R^{11}},
\\[3mm]
   \Mcb(t)
&=&
\frac{\N_2(t)- J_R^{22}K_{cb}(t) \,\mbox{\footnotesize sign}\left(\omega_1(t)  - \Rb\,\omega_2(t) \right)}{J_R^{21}}.
\end{array}
\end{equation}

\subsubsection{Full lock condition}

The full lock condition is obtained if
$\R\tras\dot\bomega(t)=0.$
Substituting \eqref{c_MA_wsq_k} in the latter yields:
\begin{equation} \label{c_MA_w_k_Full_lock}
\R\tras\J\muno\left[\M(t)
+
\R \M_c(t) \right]=0.
\end{equation}
Solving \eqref{c_MA_w_k_Full_lock} with respect to $\M_c(t)$ yields:
\begin{equation} \label{c_MA_w_k_Full_lock_condition}
\M_c(t)
=
-(\R\tras\J\muno\R)\muno\R\tras
\J\muno\M(t).
\end{equation}
Since matrix $\R$ is invertible, the full lock condition \eqref{c_MA_w_k_Full_lock_condition} can be rewritten as
$\M_c(t)
=
-\R\muno
\M(t).$

\begin{figure}[t]
\centering
\psfrag{Input torques M1, M2}[cb][cb][0.8]{Input torques $M_1(k)$ and  $M_2(k)$}
\psfrag{Input torques [Nm]}[cb][cb][0.8]{[Nm]}
\psfrag{Time [s]}[ct][ct][0.8]{Time [s]}\includegraphics[clip,width=0.88\linewidth]{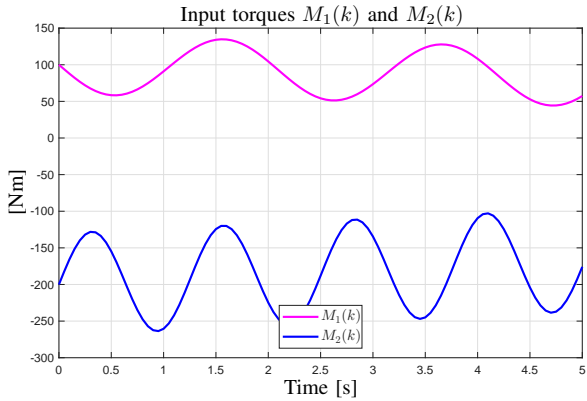}
\vspace{-1.2mm}
\caption{Input torques $M_1(k)$ and $M_2(k)$.}\label{Discrete_Figure_2_M1k_M2k}
\vspace{-4.88mm}
 \end{figure}

\section{Clutches State Control Logic}\label{control_sect}
This section addresses the description of the adopted algorithm for determining the states of the clutches in the current $k$-th time instant. The proposed control logic is reported in the pseudocode of Algorithm~1. The algorithm takes as inputs the vector $\bomega(k-1)$ containing the previous values of the angular speeds $\omega_1(k-1)$ and $\omega_2(k-1)$ of the input and output inertias $J_1$ and $J_2$, the vector $\M(k)$ containing the current values of the input torques $M_1(k)$ and $M_2(k)$ applied to the input and output inertias $J_1$ and $J_2$, and the saturation values $K_{ca}(k)$ and $K_{cb}(k)$ for the torques $\Mca$ and $\Mcb$ of clutches ``a'' and ``b'', respectively. The algorithm returns as outputs the vector $\M_c(k)$ containing the current values of the torques $\Mca$ and $\Mcb$ of clutches ``a'' and ``b'', respectively. The algorithm is conceptually organized in two phases. Firstly, the states $Lock_{ca}$ and $Lock_{cb}$ of the two clutches ``a'' and ``b'' are determined, by verifying whether the lock condition is possible for either of the two clutches - see lines from 2 to 12.
Subsequently, in the second phase, the clutches torque vector $\M_c(k)$ is computed as a function of the clutches states $Lock_{ca}$ and $Lock_{cb}$ - see lines from 13 to 28.

\section{Simulations results}\label{Simulations_results}

This section deals with the simulation of the two-speed powershift of Fig.~\ref{Sincronizzatore} using: i) the continuous-time model and ii) the discrete-time model equipped with the clutches logic in Algorithm~1. The adopted model parameters are: $J_1\!=\!0.1$ km $\mbox{m}^2\!$, $J_2\!=\!0.5$ km $\mbox{m}^2\!$, $\Ra\!=\!3$ and  $\Rb\!=\!2$. The initial conditions are $\omega_1(0)=1000$ rpm and $\omega_2(0)=100$ rpm, and the input torques $M_1(k)$ and $M_2(k)$ are shown in Fig.~\ref{Discrete_Figure_2_M1k_M2k}. The results are shown and discussed in Sec.~\ref{Simulations_sect_discrete} and Sec.~\ref{Simulations_sect_continuous} for the discrete-time case - using the step size $T_s=20$ ms - and for the continuous-time case, respectively. All simulations have been executed in the Simulink environment using MATLAB R2025a. The continuous-time simulation has been performed using the automatic solver selection option in order to let MATLAB decide the best solver, which resulted to be ode45. The comparison of the results between the discrete-time and the continuous-time cases is addressed in Sec.~\ref{Simulations_comparison}, together with the results discussion.

\begin{figure}[t]
\centering
\psfrag{Angular Speeds W1, W2}[cb][cb][0.8]{Angular speeds $\omega_2(k)$, $\omega_1(k)/R_a$ and $\omega_1(k)/R_b$}
\psfrag{Angular Speeds [rpm]}[cb][cb][0.8]{[rpm]}
\psfrag{Time [s]}[ct][ct][0.8]{Time [s]}
\includegraphics[clip,width=0.88\linewidth]{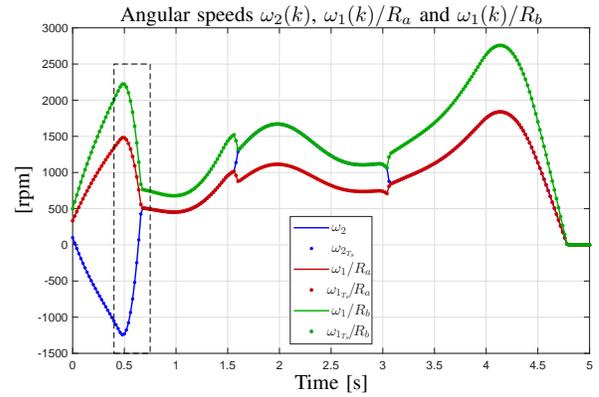}
\vspace{-1.2mm}
\caption{Discrete-time simulation: angular speeds.}\label{Discrete_Figure_3_W1k_W2k}
\vspace{-4.88mm}
 \end{figure}
%
\begin{figure}[t]
\centering
\psfrag{Torques Mca Mcb}[cb][cb][0.8]{Torques $\Mca(k)$, $\Mcb(k)$, $K_{ca}(k)$ and $K_{cb}(k)$}
\psfrag{Torques [Nm]}[cb][cb][0.8]{[Nm]}
\psfrag{Time [s]}[ct][ct][0.8]{Time [s]}
\includegraphics[clip,width=0.88\linewidth]{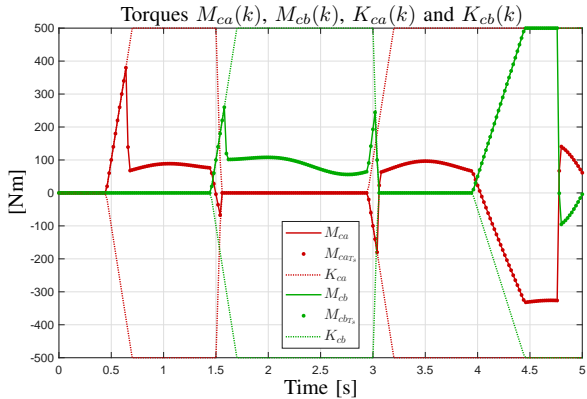}
\vspace{-1.2mm}
\caption{Discrete-time simulation: clutches torques.}\label{Discrete_Figure_4_Mca_Mcb}
\vspace{-4.88mm}
 \end{figure}

\subsection{Simulations: discrete-time  case}\label{Simulations_sect_discrete}

The angular speed
$\omega_2(k)$ of the output shaft and the angular speeds $\omega_1(k)/R_a$ and  $\omega_2(k)/R_a$ of the input shaft reported on the output shaft through the gear ratios $\Ra$ and $\Rb$
are shown in Fig.~\ref{Discrete_Figure_3_W1k_W2k}.
 The clutches torques $\Mca(k)$, $\Mcb(k)$ and their maximum values $K_{ca}(k)$, $K_{cb}(k)$  are shown in Fig.~\ref{Discrete_Figure_4_Mca_Mcb}. The dots present on the lines of Fig.~\ref{Discrete_Figure_3_W1k_W2k}
 and Fig.~\ref{Discrete_Figure_4_Mca_Mcb}
 denote the sampling instants corresponding to the step size  $T_s=20$ ms of the considered simulation, and are indeed equally spaced in time as expected.
 Fig.~\ref{Discrete_Figure_3_W1k_W2k} and Fig.~\ref{Discrete_Figure_4_Mca_Mcb} show that: 1) at $t\simeq 0.45$ gear 1 starts to be engaged (closing phase of clutch ``a''); 2) for $0.652\leq t\leq 1.543$ gear 1 is engaged (clutch ``a'' closed);  3) for $1.544\leq t\leq 1.59$  gear 1 is  disengaged and gear 2 starts to be engaged (opening clutch ``a'' and closing clutch ``b''); 4)  for $1.59\leq t\leq 3.025$ gear 2 is engaged (clutch ``b'' closed); 5)
    for $3.025\leq t\leq 3.055$  gear 2 is  disengaged and gear 1 starts to be engaged (opening clutch ``b'' and closing clutch ``a''); 6) for $3.055\leq t\leq 3.95$   gear 1 is engaged (clutch ``a'' closed); 7) for $3.95\leq t\leq 4.77$   gear 1 is still engaged
    (clutch ``a'' still closed); 8) for $t\geq 4.77$ both the gears are engaged (clutches ``a'' and ``b'' closed) corresponding to the full lock condition.


\subsection{Simulations: continuous-time  case}\label{Simulations_sect_continuous}

\begin{figure}[t]
\centering
\psfrag{Angular Speeds W1, W2}[cb][cb][0.8]{Angular speeds $\omega_2(t)$, $\omega_2(t)/R_a$ and $\omega_2(t)/R_b$}
\psfrag{Angular Speeds [rpm]}[cb][cb][0.8]{[rpm]}
\psfrag{Time [s]}[ct][ct][0.8]{Time [s]}
\includegraphics[clip,width=0.88\linewidth]{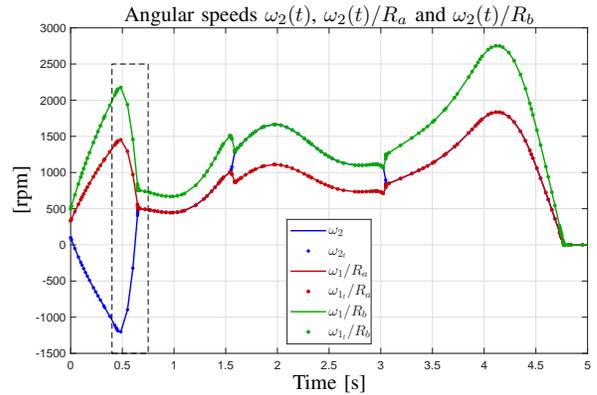}
\vspace{-1.2mm}
\caption{Continuous-time  simulation:  angular speeds.}\label{Continuous_Figure_3}
\vspace{-3.6mm}
 \end{figure}

\begin{figure}[t]
\vspace{4mm}
\centering
\psfrag{Torques Mca Mcb}[cb][cb][0.8]{Torques $\Mca(k)$, $\Mcb(k)$, $K_{ca}(k)$ and $K_{cb}(k)$}
\psfrag{Torques [Nm]}[cb][cb][0.8]{[Nm]}
\psfrag{Time [s]}[ct][ct][0.8]{Time [s]}
\includegraphics[clip,width=0.88\linewidth]{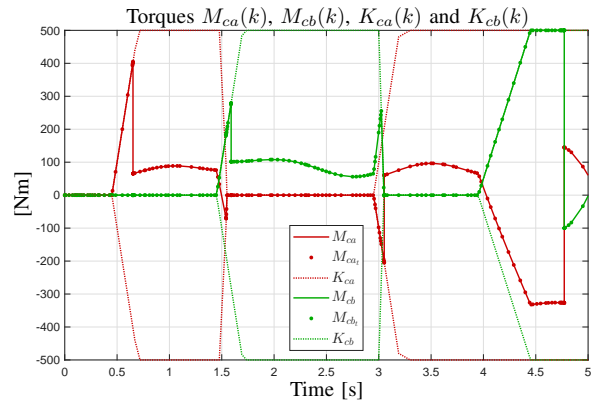}
\vspace{-1.2mm}
\caption{Continuous-time simulation: clutches torques.}\label{Continuous_Figure_4}
\vspace{-4.88mm}
 \end{figure}

The angular speeds
$\omega_2(t)$
and
$\omega_1(t)/R_a$,
$\omega_2(t)/R_a$
are shown in Fig.~\ref{Continuous_Figure_3}.
 The clutch torques $\Mca(t)$, $\Mcb(t)$ and their maximum values $K_{ca}(t)$, $K_{cb}(t)$  are shown in Fig.~\ref{Continuous_Figure_4}.
 By comparing Fig.~\ref{Continuous_Figure_3} with Fig.~\ref{Discrete_Figure_3_W1k_W2k}
 and Fig.~\ref{Continuous_Figure_4} with Fig.~\ref{Discrete_Figure_4_Mca_Mcb},
 it can indeed be seen that the results obtained using the discrete-time and continuous-time models exhibit quite a good match.
Therefore, the same observations in terms of closing and opening phases of clutches ``a'' and ``b'' as those reported in Sec.~\ref{Simulations_sect_discrete} apply in this case as well.

 However, in the continuous-time case, the dots present on the lines of Fig.~\ref{Continuous_Figure_3} and Fig.~\ref{Continuous_Figure_4},
 denoting the sampling instants corresponding to the execution of the continuous-time model using the best solver selected by MATLAB,
 {\it are not} equally spaced in time, as MATLAB automatically reduces the step size whenever needed to obtain accurate simulation results. This aspect is one of the main reasons which prevent the use of the continuous-time model for real-time execution, on the contrary with respect to the proposed discrete-time powershift model. Specifically, a more detailed discussion of this aspect can be found in the next Sec.~\ref{Simulations_comparison}.

%

%
 %

\begin{figure}[t]
\centering
\vspace{2mm}
\psfrag{Angular Speeds W1, W2}[cb][cb][0.8]{Angular speeds $\omega_2(t)$, $\omega_2(t)/R_a$ and $\omega_2(t)/R_b$}
\psfrag{Tictoc}[cb][cb][0.8]{Execution Frequency}
\psfrag{Angular Speeds [rpm]}[cb][cb][0.8]{[rpm]}
\psfrag{diff(t)}[cb][cb][0.8]{[Hz]}
\psfrag{Time [s]}[ct][ct][0.8]{Time [s]}
\includegraphics[clip,width=0.88\linewidth]{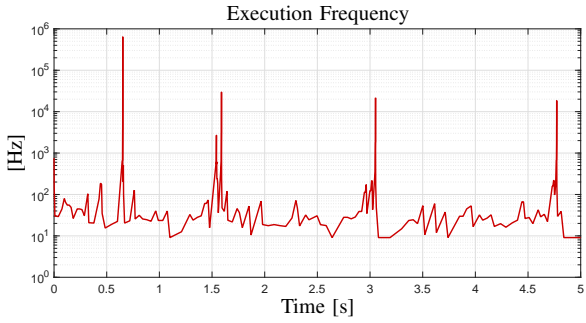}
\vspace{-1.2mm}
\caption{Continuous time  case: execution frequency, denoting how frequent the execution steps of the solver algorithm are versus the simulation time.}\label{Continuous_Figure_5}
\vspace{-3.6mm}
 \end{figure}

\subsection{Continuous-Time vs Discrete-Time Models: Results Comparison and Discussion}\label{Simulations_comparison}

The analysis of Fig.~\ref{Continuous_Figure_3} and Fig.~\ref{Continuous_Figure_4} in the previous section has revealed that the dots denoting the sampling instants corresponding to the execution of the continuous-time model are indeed not equally spaced in time.
This aspect can be clearly seen from Fig.~\ref{Continuous_Figure_5}, showing the execution frequency in a logarithmic scale, which denotes how frequent the execution steps of the solver algorithm are as a function of the simulation time.

\begin{figure}[t]
\vspace{4mm}
\centering
\psfrag{Angular Speeds W1, W2}[cb][cb][0.8]{Angular speeds $\omega_2(t)$, $\omega_2(t)/R_a$ and $\omega_2(t)/R_b$}
\psfrag{Angular Speeds [rpm]}[cb][cb][0.8]{[rpm]}
\psfrag{Time [s]}[ct][ct][0.8]{Time [s]}
\includegraphics[clip,width=0.82\linewidth]{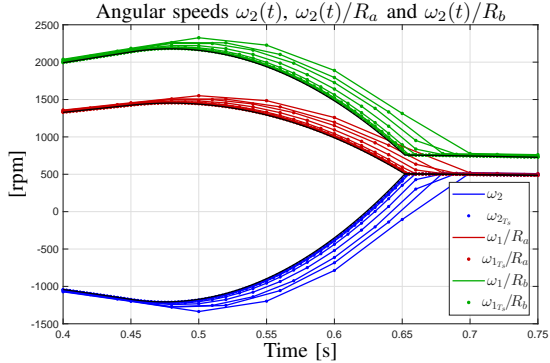}
\vspace{-2.2mm}
\caption{Zoom-in of the dashed rectangular zone in Fig.~\ref{Discrete_Figure_3_W1k_W2k} and Fig.~\ref{Continuous_Figure_3}, showing a comparison between the continuous-time case (black lines) and the discrete-time case (colored lines) when the step size $T_s$ takes the following values: $T_s\in\{0.05\, 0.04,\, 0.03,\, 0.02,\, 0.01,\, 0.005,\, 0.0025\}$.}\label{Discrete_Figure_3_W1k_W2k_Confronta}
\vspace{-3.68mm}
 \end{figure}

By comparing Fig.~\ref{Continuous_Figure_5} with Fig.~\ref{Continuous_Figure_4} and Fig.~\ref{Continuous_Figure_3},
 it is possible to appreciate that peaks of the execution frequency occur in correspondence of the simulation phases associated with the clutches engagement and disengagement, as these are the most critical ones from a simulation point of view. However, such densification of execution steps is not allowed in real-time implementation. Furthermore, Fig.~\ref{Discrete_Figure_3_W1k_W2k_Confronta} shows that the proposed discrete-time model of the two-speed powershift gives results that are very close to those of the continuous-time model. The upper boundary of the variable step size $T_s$ has been set to $10^{-3}$ for the continuous-time model in this figure, to increase its precision and obtain an accurate solution which the results of the discrete-time model can be compared against. Additionally, the results of the discrete-time model become more and more accurate as the fixed step size decreases.

\begin{figure}[t]
\centering
\includegraphics[clip,width=0.88\linewidth]{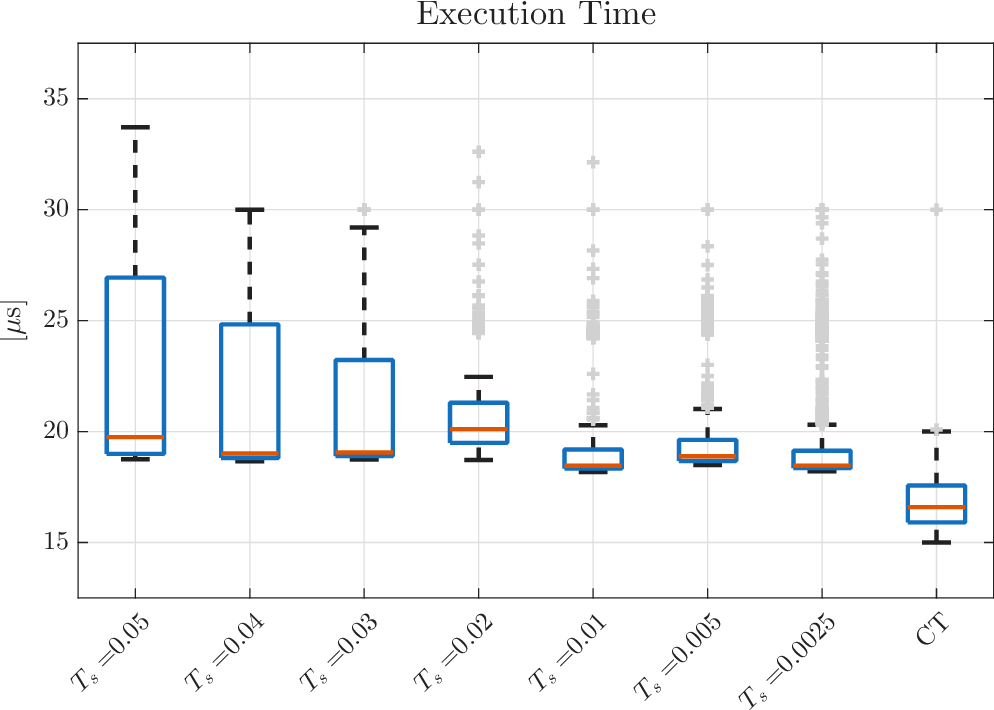}
\vspace{-3.6mm}
\caption{Average time (over $20$ executions) of the single simulation step for the continuous-time model (CT) and for the discrete-time model when $T_s\in\{0.05\, 0.04,\, 0.03,\, 0.02,\, 0.01,\, 0.005,\, 0.0025\}$.}\label{execution_time_boxplots}
\vspace{-3.68mm}
 \end{figure}

Finally, the boxplots of Fig.~\ref{execution_time_boxplots} show the average time (over $20$ executions) of the single simulation step for the continuous-time model and for the discrete-time model using different step sizes $T_s$. Here, the orange hotizontal lines denote the median values, the light-blue rectangles denote the interquartile range spanning from the first to the third quartiles, while the grey crosses denote outliers. From Fig.~\ref{execution_time_boxplots}, it can be seen that the area of the interquartile range rectangles tends to decrease as the step size $T_s$ decreases.
Furthermore, it can be noted that the average execution time of the single simulation step for the continuous-time model and for the discrete-time model (using different values of $T_s$) are comparable, with negligible differences considering that the y-axis scale is expressed in $\mu$s.

The obtained simulation results show that the proposed discrete-time model of the two-speed powershift equipped with the clutches logic of Algorithm~1 give comparable performances - both in terms of average execution time and in terms of simulation results - as those obtained using the continuous-time model. At the same time, on the contrary with respect to the continuous-time model adopting a variable-step solver, the proposed discrete-time model equipped with the proposed clutches logic is suitable for real-time simulation.

 %

\section{Conclusions}\label{conclusions_sect}

This paper has addressed the proposal of a new discrete-time approach to model the clutches engagement/disengagement in a two-speed powershift, allowing to account for both the cases of single clutch engagement to perform gearshift and of simultaneous clutch engagement to simulate the full lock condition. Furthermore, a control logic for the clutches
engagement and disengagement phases has also been developed.
While the advantages of the proposed approach have been shown through extensive simulation results with reference to the proposed two-speed powershift case study, it is worth noting that the proposed work shows great potential for future extension to the modeling of several other physical systems commonly found in vehicle powertrains such as, among others, torque converters and brakes.

\addtolength{\textheight}{-12cm}

\bibliographystyle{IEEEtran}

\bibliography{Discrete_Two_Speeds_Power_Shift}

\end{document}